\documentclass{article}

\usepackage[english]{babel}

\usepackage{graphics}
\usepackage{graphicx}
\usepackage{multirow}
\usepackage{caption}
\usepackage{mathtools}
\usepackage{extarrows}
\usepackage{float}

\usepackage[numbers,super,sort&compress]{natbib}

\usepackage[
    letterpaper,
    top=1.5cm,
    bottom=1.5cm,
    left=1.5cm,
    right=1.5cm,
    marginparwidth=1.5cm
]{geometry}

\usepackage{authblk}

\usepackage[hyphens]{url}
\usepackage{xurl}
\usepackage[
    colorlinks=false,
    breaklinks=true
]{hyperref}
\usepackage{lineno}
\usepackage{xcolor}
\usepackage{abstract}

\date{}

\providecommand{\keywords}[1]
{
    \fontsize{12}{14}\selectfont
    \textbf{Keywords:} #1
}

\title{
\textbf{
\textsf{
Computational Insights into Mechanostability and Dissociation Dynamics of the Dengue Virus Envelope Protein Ectodomain Dimer Across pH and Temperature Gradients
}}
}

\author[1]{\fontsize{12}{14}\selectfont Georcki Ropón-Palacios}
\author[1]{\fontsize{12}{14}\selectfont Luís G. F. Crespi}
\author[2]{\fontsize{12}{14}\selectfont Jakub Rydzewski}
\author[3]{\fontsize{12}{14}\selectfont Walter Rocchia}
\author[1]{\fontsize{12}{14}\selectfont Alexandre S. de Araujo$^{*}$}

\affil[1]{\fontsize{10}{14}\selectfont
Department of Physics, São Paulo State University (UNESP),
Institute of Biosciences, Humanities and Exact Sciences,
São José do Rio Preto, SP, 15054-000, Brazil
}

\affil[2]{\fontsize{10}{14}\selectfont
Institute of Physics, Faculty of Physics, Astronomy and Informatics,
Nicolaus Copernicus University in Torun,
ul. Grudziadzka 5, Toruń 87-100, Poland
}

\affil[3]{\fontsize{10}{14}\selectfont
CONCEPT Lab, Istituto Italiano di Tecnologia (IIT),
Via Enrico Melen, 83 -- 16152 Genova, Italy
}

\affil[*]{\fontsize{10}{14}\selectfont
Corresponding author:
}

\affil[]{\fontsize{10}{14}\selectfont
E-mail: alexandre.suman@unesp.br \\
Phone: +55 (17) 3221-2566
}

\begin{document}
\maketitle
\newpage
\begin{abstract}
\fontsize{12}{14}\selectfont

Dengue virus (DENV) is an enveloped flavivirus of major public health importance. Its envelope (E) protein mediates viral entry through homodimer dissociation and subsequent membrane fusion, making it one of the principal targets for antiviral strategies. To investigate the molecular determinants of this process, we performed steered molecular dynamics (SMD) simulations of the E protein ectodomain (ecE) dimer from DENV-2 and DENV-3 under variable pH and temperature conditions. Force-extension analyses revealed a highly stable interface for both serotypes, with rupture forces exceeding 1000 pN. pH and temperature had modest effects on overall mechanical resistance. However, DENV-3 displayed a distinct sensitivity to thermal stress compared to DENV-2. We identified a robust, asymmetric dissociation pathway across all conditions, characterized by a metastable intermediate state involving partial dimer opening and exposure of the fusion loop (FL). This intermediate exposes an immunodominant epitope and persists under physiological conditions, suggesting it as a viable target for therapeutic intervention. The primary contributions to the interfacial interaction network were found to arise from van der Waals interactions, followed by hydrogen bonds, salt bridges, and pi-cation interactions. DENV-3 exhibited a slightly greater contribution from polar interactions involving domains EDI, EDII, and EDIII. Furthermore, pairwise occupancy analysis identified pH-sensitive contacts that are disrupted under acidic conditions, particularly in DENV-3, providing mechanistic insight into the early stages of ecE dissociation. Together, these findings provide structural insights into the dissociation mechanism, identifying key metastable states and pH-sensitive interactions that could be exploited to develop antivirals that stabilize the dimer and prevent viral fusion.

\end{abstract}


\newpage

\section{INTRODUCTION}\label{s:introduction}
\fontsize{12}{14}\selectfont
Dengue virus (DENV) is a mosquito-borne pathogen prevalent in tropical and subtropical regions worldwide, estimated to cause approximately 390 million infections annually, including about 96 million clinically apparent cases \cite{A6, A9}. Annual dengue-associated deaths are generally estimated to exceed 20,000 \cite{A10, A11}, with clinical manifestations ranging from mild febrile illness to severe dengue hemorrhagic fever and dengue shock syndrome, further amplified by antibody-dependent enhancement across its four antigenically distinct serotypes \cite{A4, A11, A14, A18}. The infection cycle begins when DENV binds to receptors on target cells and is internalized via clathrin-mediated endocytosis, whereupon progressive acidification of the endosomal lumen triggers structural rearrangements in the viral surface proteins that drive membrane fusion and release of the viral genome into the host cytoplasm \cite{A12, A13, A14, A16, A17}. Central to this process is the E protein, whose conformational plasticity governs each step from cell recognition to membrane fusion, emerging as the primary target for neutralizing antibodies and antiviral intervention.
The E protein is a class II fusion transmembrane protein packed as homodimers on the virion surface  \cite{A12}, differing by 30–35\% in amino acid sequence across the four DENV serotypes \cite{A8} while maintaining an invariant global fold. Its ectodomain (ecE) is organized into three distinct domains (EDI–EDIII): domain II (EDII) harbors the hydrophobic fusion loop (FL) (residues 98–110), which is highly conserved across all flaviviruses, and, under acidic conditions in the endosome, ecE dimers dissociate \cite{A14, A15, A23}, allowing the monomers to extend their fusion loops to interact with and anchor to the host cell membrane \cite{A13, A17, A23}. This process depends on highly conserved fusion loop residues, particularly W101, L107, and F108, which insert into the target membrane and help form a hydrophobic anchor in the lipid bilayer. K110 is also important for infectivity; structural studies of the post-fusion E trimer suggest that this residue may contribute to stabilization of the trimer apex by coordinating a chloride ion near the three-fold symmetry axis of the trimer \cite{A16, A23, A27}. EDII also carries the N67 glycosylation site, which contributes to host cell receptor recognition \cite{A16, A19}. Domain III (EDIII) constitutes the primary site for receptor binding, connecting the ectodomain to the stem region. Anchored to the outer leaflet of the viral membrane, the stem region consists of two $\alpha$-helices that rearrange during the fusion process. Finally, the C-terminal transmembrane domain comprises two antiparallel coiled-coils (TM1 and TM2) that stabilize the fusogenic trimer and complete membrane fusion with the endosomal bilayer \cite{A16}. Together, this modular architecture underlies both the functional versatility and the intrinsic metastability of the ecE homodimer.
The structural organization of the E protein described above is, however, metastable by nature \cite{A22}: upon acidification of the endosomal lumen (pH $\sim$5.0–6.5) \cite{A14}, the homodimeric arrangement is destabilized and the dimers dissociate, allowing the monomers to reorient and insert their fusion loops into the host membrane, triggering a conformational transition toward the fusogenic homotrimer \cite{A7, A12, A14, A15, A28}. This pH-driven process is primarily mediated by key histidine residues (e.g., H144, H244, H282, and H317, numbered in DENV-2) and, as recently reported, by an EDI–EDIII interface-specific cluster of amino acids (e.g., N-terminus R9, D42, H144, and E368) that coordinate the conformational rearrangements required for trimer formation \cite{A14}. In this trimeric configuration, DIII of the E protein folds back on itself toward the already anchored fusion loop, promoting the final fusion between the endosomal and viral membranes \cite{A14, A16, A23}. Despite being the primary trigger of this transition, pH does not act in isolation. Temperature has emerged as a relevant physicochemical factor in the destabilization of the ecE dimeric interface, and has further been shown to favor a phenomenon known as viral breathing at the virion level \cite{A24, A20, A25}, which drives morphological changes in the virion. Studies by Kudlacek et al. (2018) \cite{A20} demonstrated that increasing temperature from 23 $^\circ\mathrm{C}$ to 37 $^\circ\mathrm{C}$ weakens this interface, shifting the dimer–monomer equilibrium in solution toward the monomeric form; notably, DENV-2 exhibits a steep temperature dependence, whereas DENV-3 is inherently less stable and exists primarily as a monomer even at lower temperatures \cite{A20}. This temperature-induced tendency toward monomerization has direct implications for the development of therapeutics, particularly antibodies that recognize quaternary epitopes dependent on the integrity of the dimer, whose neutralizing potential is compromised at physiological temperature \cite{A21, A20}.
Taken together, the available evidence highlights the vulnerability of quaternary epitope-targeting strategies under physiological conditions and points to the dimeric interface itself as a more tractable therapeutic target. Molecular dynamics simulations offer atomic-resolution insight into the conformational intermediates populated during dimer dissociation under controlled pH and temperature conditions, complementing experimental structural data. Indeed, in previous work we identified a metastable state designated the “compensatory embrace” \cite{roponpalacios2026}, in which the FL of one monomer becomes exposed while that of the other remains occluded, suggesting that the dissociation pathway could be populated by key intermediate configurations with potential therapeutic relevance. This observation raises the possibility that a more systematic exploration of the dissociation process may uncover additional metastable states amenable to therapeutic intervention, whether through stabilization of the pre-fusion dimer or disruption of specific interfacial interactions.
Motivated by these observations, this study employs steered molecular dynamics (SMD) simulations (Fig. 1) to characterize the mechanostability and dissociation pathways of the DENV ecE homodimer under varying pH and temperature conditions, providing structural insights that may guide the rational design of dimer-stabilizing therapeutics.

 \begin{figure}[H]
        \includegraphics[width=1\textwidth]{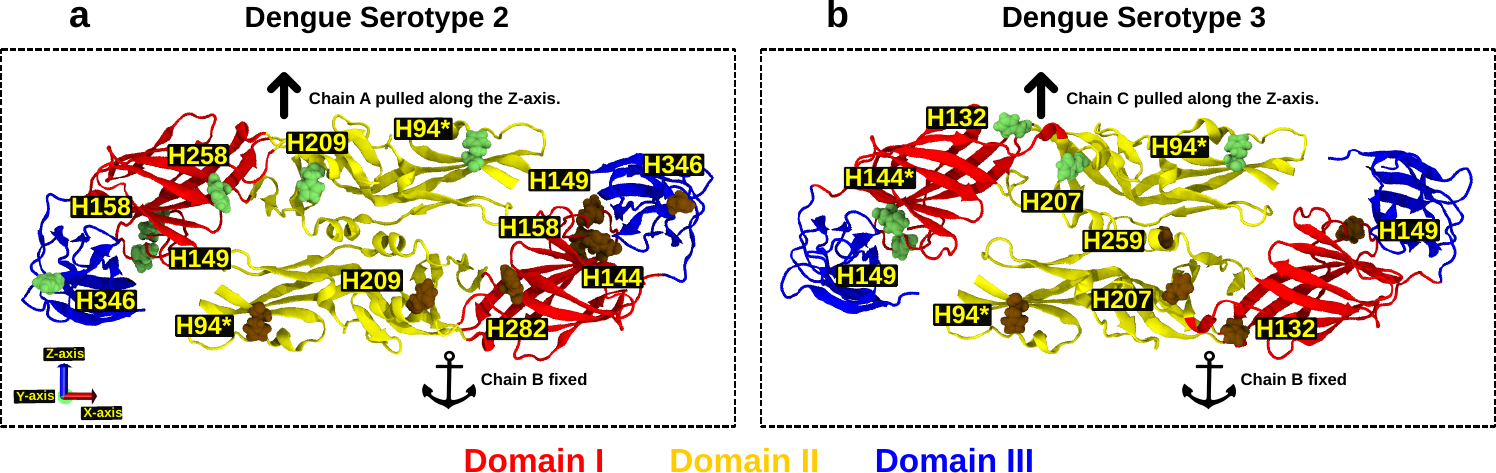}
    \centering
    \caption{\textbf{SMD setup and histidine protonation states of DENV-2 and DENV-3 ecE dimers.} (a,b) Structural models showing the pulled and fixed monomers. Highlighted spheres denote protonated histidines under acidic conditions. All labeled residues are protonated at pH 5, whereas only those marked with an asterisk remain protonated at pH 6; all histidines are unprotonated at pH 7.
    \label{fig:Fig_1}}
\end{figure}

\section{MATERIALS AND METHODS}

\subsection*{Structural modeling and system preparation}

The atomic coordinates of dengue virus E protein ectodomain (ecE) dimers (serotypes DENV-2 and DENV-3; PDB IDs 1OAN and 7A3S, respectively) were retrieved from the Protein Data Bank \cite{M1, M2}. Histidine protonation states at pH 5, 6, and 7 were assigned using PROPKA3 \cite{M3} through the PDB Reader \& Manipulator module of CHARMM-GUI \cite{M4, M5}, which was also used to reconstruct missing loops and unresolved atoms. All-atom topologies (PSF) and coordinates (PDB) were generated using the CHARMM36m force field with CMAP corrections \cite{M6, M7}.
System setup from this point forward was performed using the MSTBx v0.5 toolkit (https://github.com/SSiLiq/MSTBx). Prior to solvation, each ecE dimer was oriented to align the pulling direction along the Z-axis, defined by the C$\alpha$ atoms of residue 256 from each monomer for DENV-2, and residues 215 and 213 from the respective monomers for DENV-3. The oriented dimer was then solvated in a TIP3P water box with a base padding of 30 \text{\AA} in the xy-plane and 15 \text{\AA} along the z-axis, relative to the solvent-accessible surface, to prevent self-interaction. To accommodate monomer displacement during the SMD simulations, an additional 50 \text{\AA} of box space was added exclusively along the positive z-axis. This extension resulted in final box dimensions approximately 199 $\times$ 199 $\times$ 234 \text{\AA} for DENV-2 and 171 $\times$ 171 $\times$ 206 \text{\AA} for DENV-3. Finally, to reproduce physiological conditions, Na$^{+}$ and Cl$^{-}$ ions were added to neutralize the systems and achieve an ionic strength of 150 mM.

\subsection*{Steered molecular dynamics simulations}

All simulations were performed with NAMD v3.0b6 \cite{M8, M9} using a common equilibration and SMD protocol applied to all systems. A 2 fs integration timestep was employed during the equilibration and SMD production stages. The following nonbonded and electrostatic settings were applied consistently across minimization, equilibration, and SMD: long-range electrostatics and nonbonded interactions were both evaluated at every step; all covalent bonds involving hydrogens were constrained; long-range electrostatics were computed using the particle mesh Ewald method with a 1 \text{\AA} grid spacing \cite{M10, M11}; and short-range vdW interactions were truncated at 12 \text{\AA} with force switching enabled and a switching onset at 10 \text{\AA}. For the dynamics stages, temperature and pressure were controlled with a Langevin thermostat (damping coefficient of 1 ps$^{-1}$) \cite{M12} and a Nosé–Hoover Langevin piston barostat (oscillation and decay periods of 200 fs and 100 fs, respectively) \cite{M13, M14}.
Each serotype (DENV-2 and DENV-3) was simulated under a full combination of three pH conditions (5, 6, and 7) and three temperatures (301, 310, and 313 K; approximately 28, 37, and 40 $^\circ\mathrm{C}$), yielding nine conditions per serotype and 18 systems in total. Each system was first subjected to 50,000 steps of energy minimization using the conjugate gradient algorithm. Equilibration was then carried out in two successive stages: (1) 2 ns of NVT equilibration at the target temperature; and (2) 5 ns of NPT equilibration at the same target temperature and 1.01325 bar. Positional restraints of 5 kcal mol$^{-1}$ \text{\AA}$^{-2}$ were applied to the backbone atoms of the dimer throughout minimization and both equilibration stages.
Following equilibration, SMD simulations were conducted for 8 ns per system in the NPT ensemble using the Colvars module \cite{M15}. The force applied to the target atoms was calculated according to:
\[
F_z = -k\left[\xi(t)-\left(\xi_0+vt\right)\right],
\qquad
v = \frac{\xi_{\mathrm{target}}-\xi_0}
{N_{\mathrm{steps}}\,\Delta t}
\]
where $\xi$($t$) is the instantaneous value of the collective variable, $k$ is the spring force constant (1.5 kcal mol$^{-1}$ \text{\AA}$^{2}$), $\xi_0$ is the initial restraint center, defined as the initial Z-coordinate of the C$\alpha$ center of mass of the pulled monomer relative to the dummy atom at the origin (-18.9 \text{\AA}), and $t$ is time. The pulling velocity $v$ was not prescribed directly but followed from the linear displacement of the restraint center, where $\xi$ target is the final Z-coordinate of the restraint center (61.1 \text{\AA}), Nsteps is the total number of steering steps (targetNumSteps = 4 $\times$ 10$^{6}$), and $\Delta$$t$ is the integration timestep (2 fs), giving a constant velocity of 10 \text{\AA} ns$^{-1}$. The collective variable was defined through the Colvars module as a distance Z variable, tracking the projection onto the Z axis of the center of mass calculated from all C$\alpha$ atoms of the pulled monomer relative to a fixed dummy atom placed at the origin, rather than the center-of-mass distance between the two monomers. Over the full steering protocol, the restraint center was displaced linearly along the Z axis from -18.9 \text{\AA} to 61.1 \text{\AA}, corresponding to a total displacement of 80 \text{\AA}. For both serotypes, Monomer B served as the anchor, with a harmonic restraint (force constant of 5 kcal mol$^{-1}$ \text{\AA}$^{-2}$) applied to the center of mass of its C$\alpha$ atoms, while the moving restraint was applied to the center of mass of the C$\alpha$  atoms of the steered monomer (Monomer with chainID A for DENV-2 and Monomer with chainID C for DENV-3). Applied force, restraint center, and accumulated work were recorded every 0.2 ps, while atomic coordinates were saved every 10 ps, for all conditions and temperatures.

\subsection*{Trajectory data analysis}

Mechanical and structural dimer dissociation profiles were characterized from SMD simulations performed at 28, 37, and 40 $^\circ\mathrm{C}$ under pH 5, 6, and 7 conditions. Force–extension data were extracted directly from the Colvars module. Trajectory processing and structural analyses were carried out using VMD 2.0.0b1 \cite{M16} and MDAnalysis 2.7 \cite{M17}, complemented by custom Python, Tcl, and Bash scripts. All analysis scripts are publicly available at https://github.com/SSiLiq/Georcki-CompBiology-Biophysics.
The final rupture force for each replica was determined using a peak detection algorithm applied to force profiles previously smoothed by a 50-frame centered moving average. The rupture event was defined as the last peak identified by the SciPy \texttt{find\_peaks} function, using a minimum prominence threshold of 2.0 to discard minor fluctuations.
Intermonomeric contacts were quantified considering only C$\alpha$ atoms located at the dimer interface and adopting a distance cutoff of 8.0 \text{\AA}. Residue-level interaction patterns were further characterized using GetContacts (available in: https://getcontacts.github.io/), enabling the identification and classification of specific noncovalent interactions throughout the dissociation process.
Conformational dynamics and dimer orientation were characterized by monitoring the Euclidean distance between the centers of mass of the C$\alpha$ atoms belonging to the cradle region (residues 310–320 of chain B) and the FL region (residues 98–110 of the pulled monomer; chain A in DENV-2 and chain C in DENV-3). The tilt angle ($\theta$), describing the global orientation of the pulled monomer relative to the pulling (Z) axis, was calculated from the dominant principal axis of the entire pulled monomer. To obtain geometry-based principal axes, all atoms were assigned unit masses during the inertia tensor calculation. The tilt angle was defined from Equation 2, where pz corresponds to the Z-component of the dominant principal-axis unit vector, restricting angular values to the first quadrant (0–90°).
\[
\theta = \arccos\left(\left|p_z\right|\right)
\]
To identify representative conformational states along the dissociation pathway, three reference states were defined for each system: (i) the initial state, corresponding to the average distance from the center of mass of Cradle-FL and tilt angle of the first frame from all replicas; (ii) an intermediate state, defined as the average on the fifth occurrence in which the cradle–FL distance reached or exceeded 12 \text{\AA} in each replica, ensuring stability and robustness; and (iii) the final state, corresponding to the average of the last frame from all replicas.

\section{RESULTS}

\subsection*{Assessing whether pH and temperature perturb the mechanostability of the DENV ecE} 
The role of pH in dimer dissociation is well established, promoting large conformational rearrangements \cite{A14, A7, A12, A15}. Temperature also proves relevant to the dimer–monomer equilibrium. At physiological conditions (37 $^\circ\mathrm{C}$), DENV-2 shifts toward the monomeric form. DENV-3, in contrast, displays greater thermal susceptibility, beginning to adopt the monomeric state at temperatures below 37 $^\circ\mathrm{C}$ \cite{A20}. However, to our knowledge, no information exists on how mechanostability may be affected by them.
To assess this, we first analyzed the average force-extension profiles (Fig. 2a) under the different conditions. Overall, both serotypes exhibited a pronounced force peak (corresponding to the maximum force), exceeding 1000 pN. In DENV-2, following this peak, the force rapidly decreased to below 500 pN over the next $\sim$10–20 \text{\AA} of displacement along the pulling coordinate. In DENV-3, by contrast, within this same distance range the profile still displays a secondary peak above 500 pN, indicating a second rupture point. Regarding the effect of pH, the profiles were comparable in DENV-2. In DENV-3, however, the force-extension profile  differed markedly at pH 7 compared to pH 5 and 6 at 28 and 37 $^\circ\mathrm{C}$: at 28 $^\circ\mathrm{C}$, the neutral medium requires a higher force for the second rupture, whereas at 37 $^\circ\mathrm{C}$ this higher force requirement for the second rupture shifts to pH 5. Conversely, at 40 $^\circ\mathrm{C}$, the profiles at pH 7 and pH 5 remained comparable, exhibiting only minor differences. With respect to temperature, in DENV-2 the profiles are comparable, although slightly higher at pH 6 and 37 $^\circ\mathrm{C}$, but no clear effect is observed. In DENV-3, however, the force-extension profile shows a slight reduction in the maximum-force peak as temperature increases, such that, at 40 $^\circ\mathrm{C}$, the maximum-force peak is similar across the three pH values.
To further understand the effect of pH and temperature on the dimer interface and its relationship with mechanostability, we analyzed the number of contacts between the C$\alpha$ atoms of the dimer interface (Fig. 2b). Overall, the number of contacts differed between DENV-2 and DENV-3, with 104.1 $\pm$ 1.6 and 95.9 $\pm$ 1.3 for all replicas of the respective viruses under all conditions, measured at frame zero of the production run, after equilibration. In addition, we found a profile related to that of the force-extension curve (Fig. 2a). The contact loss profile (Fig. 2b) corroborates the mechanical behavior evidenced by the force-extension curves (Fig. 2a). In the DENV-2 complex, a marked disruption of interactions occurs between -18 \text{\AA} and -13 \text{\AA} at pH 5 and 6 at 28 $^\circ\mathrm{C}$, resulting in 66 and 67 contacts, respectively. At pH 7 and the same temperature, this profile extends to -11 \text{\AA} with 67 contacts, followed by a brief plateau that persists up to -7 \text{\AA}, compared to -9 \text{\AA} for pH 5 and 6. At 37 $^\circ\mathrm{C}$, the contact decay profile is similar across all pH conditions, reaching a threshold at -13 \text{\AA} with roughly 65 contacts; yet, at pH 5, a sharper decline continues down to -5 \text{\AA} with 55 contacts, whereas pH 7 and 6 retain 57 and 56 contacts in this range, respectively. Finally, at 40 $^\circ\mathrm{C}$, the contact profiles remain comparable across all three pH conditions. While DENV-3 at 28 $^\circ\mathrm{C}$ shows a comparable initial decay down to -14 \text{\AA}, resulting in 63, 65, and 62 contacts at pH 7, 6, and 5, respectively, it diverges markedly in the subsequent phase. The ensuing plateau features a much gentler rate of decline that stretches up to 20 \text{\AA}, sustaining a higher contact count than DENV-2 throughout this trajectory. Comparing DENV-3 across temperatures reveals that at pH 5, the contact count is higher at 37 $^\circ\mathrm{C}$ than at 28 $^\circ\mathrm{C}$, whereas at pH 7, the count decreases at 37 $^\circ\mathrm{C}$ than at 28 $^\circ\mathrm{C}$. At 40 $^\circ\mathrm{C}$, the profiles for pH 7 and 5 remain comparable, in contrast to pH 6, which clearly preserves a distinctly lower number of contacts. Subsequent analysis of the pH effect revealed that, for DENV-2, variations in environmental acidity did not exert a significant influence on the maintenance of interactions, except at 28 $^\circ\mathrm{C}$ and pH 7, a condition that exhibited the highest number of contacts compared to the other pH values evaluated at this temperature. In contrast, DENV-3 proved to be more susceptible under specific conditions, displaying an accelerated loss of contacts at 40 $^\circ\mathrm{C}$ and pH 6; a trend also observed to a lesser extent at 37 $^\circ\mathrm{C}$ (pH 7 and 6). With respect to temperature, we did not observe a clear effect on the number of contacts in either serotype; the only more relevant difference corresponds precisely to DENV-3 at 40 $^\circ\mathrm{C}$ and pH 6, mentioned above.

 \begin{figure}[H]
        \includegraphics[width=1\textwidth]{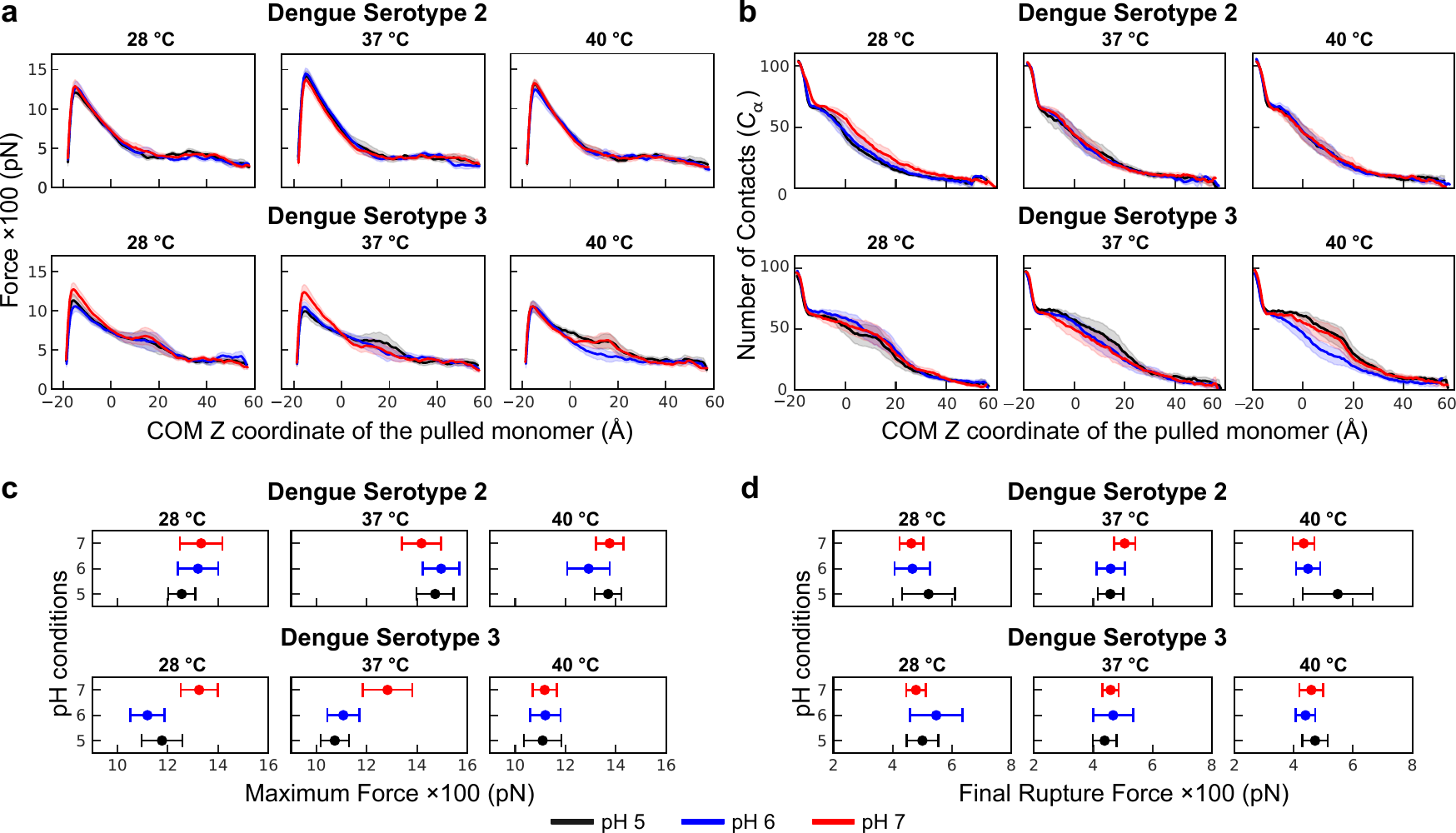}
    \centering
    \caption{\textbf{Dissociation profiles of DENV-2 and DENV-3 ecE dimers obtained from SMD simulations.} (a) Pulling force versus displacement. (b) Intermonomeric C$\alpha$ contacts. (c) Maximum pulling force. (d) Final rupture force. Data were obtained at 28, 37, and 40 °C under pH 5 (black), pH 6 (blue), and pH 7 (red) conditions. Values represent the mean of 20 independent replicas; shaded regions and error bars indicate 95\% confidence intervals.
    \label{fig:Fig_2}}
\end{figure}

To quantitatively analyze the maximum force, which corresponds to the main rupture force (Fig. 2a, around -11 \text{\AA}), under the different pH and temperature conditions, we present in Figure 2c the mean and the 95\% confidence interval, represented by the error bars. Regarding the effect of pH in DENV-2, we found that the values are comparable, showing no significant differences. In DENV-3, the maximum force at pH 7  tended to be greater than that of the other two pH values at 28 and 37 $^\circ\mathrm{C}$. Next, we analyzed the effect of temperature. In DENV-2, the maximum force increased across the three pH values when going from 28 to 37 $^\circ\mathrm{C}$, being slightly lower at pH 7 relative to the other two pH values, though without evident significance. In DENV-3, temperature appeared to have a more discernible, though not strictly monotonic, effect: the maximum force at pH 5 and pH 6 showed a decreasing trend from 28 to 37 and 40 $^\circ\mathrm{C}$, whereas at pH 7 a reduction was apparent only at 40 $^\circ\mathrm{C}$.
In addition to the maximum force, we quantitatively evaluated the final rupture force (Fig. 2d), which precedes the breaking of the remaining contacts still present at the interface. Interestingly, we observed that these remaining contacts correspond, in general and across all conditions for both serotypes, to interactions between domains EDI and EDIII of the anchored monomer and the FL of the pulled monomer, also observed in Figure 3d,e (and Figure S2a,b). Regarding the effect of pH in DENV-2, we note that there is no clear evidence of such an effect on the final rupture force, since the values are comparable and their confidence intervals overlap. The same absence of evidence was observed in DENV-3. Finally, when analyzing the effect of temperature, we reached the same result obtained for pH.

\subsection*{Mechanical unbinding of DENV ecE dimers} 

During the course of infection, dimer dissociation is one of the key steps, driven by the pH change that acidifies the late endosome ($\sim$pH 6.5 to 5.0)\cite{A14, A7}. To assess this process, we inspected the trajectories visually. Interestingly, we found that, across all conditions, the dissociation mechanism was asymmetric (see Fig. S1,S2). This asymmetry is consistent with the state preceding the final rupture force, in which the remaining contacts are located specifically between the FL and its cradle, defined here as the C$\alpha$ contacts of the opposing monomer located within 8 \text{\AA} of the FL (Fig. 3d,e).
Next, to describe this dissociation mechanism, we defined two geometric collective variables (CVs): (1) the tilt angle between the Z axis and the principal axis of the pulled monomer; and (2) the distance between the center of mass of the cradle on the anchored monomer and the FL of the pulled monomer (Fig. 3a). The 2D probability analysis of these two CVs allowed us to quantitatively identify an intermediate state consistent with the asymmetric dissociation (Fig. 3b,c and Fig. S1,S2). The probability associated with this state refers to its local density in the normalized tilt angle and cradle–FL distance maps generated from the 20 concatenated replicas, with values generally below 0.5 on the normalized color scale (Fig. 3b,c and Fig. S1). Three representative states were defined: (i) the initial state, from the mean CV values of the first frame of all replicas; (ii) the intermediate state, from the fifth occurrence in each replica at which the cradle–FL distance reached or exceeded 12 \text{\AA}, minimizing transient crossings; and (iii) the final state, from the mean CV values of the last frame of all replicas.

 \begin{figure}[H]
        \includegraphics[width=1\textwidth]{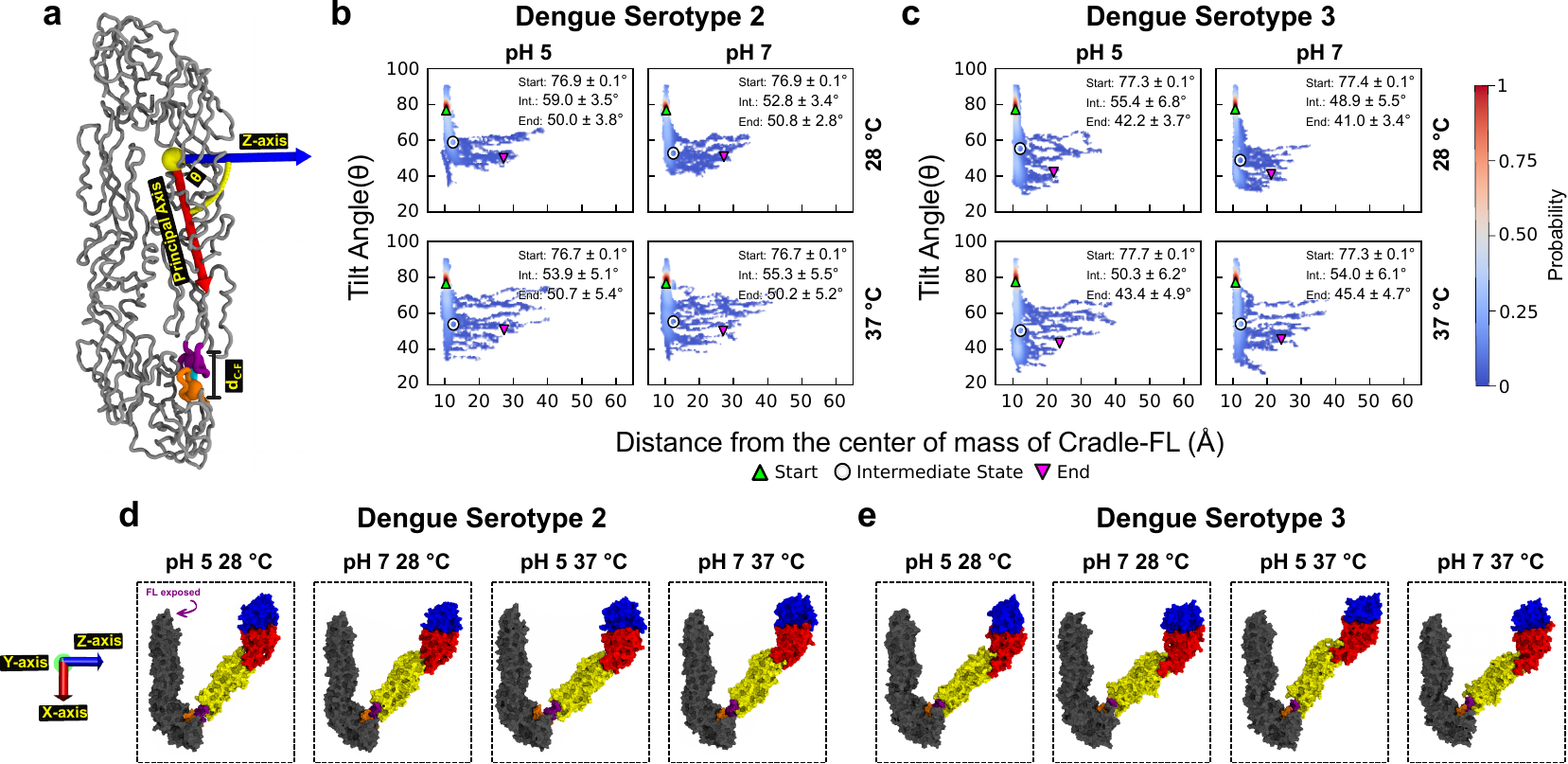}
    \centering
    \caption{\textbf{Conformational landscapes of DENV-2 and DENV-3 ecE dimers during dissociation.} (a) Definition of the reaction coordinates: monomer tilt angle ($\theta$) and cradle–FL (C-F) distance. (b,c) Normalized probability density maps showing the transition from the Start to Intermediate and End states. Symbols indicate averages from 20 independent replicas. (d,e) Representative final expanded conformations highlighting FL exposure following dimer dissociation.
    \label{fig:Fig_3}}
\end{figure}

Although the asymmetric state is present across all conditions, we next focused on the most biologically relevant ones (28/37 $^\circ\mathrm{C}$ and pH 5/7) to examine the influence of pH and temperature on these states (Fig. 3). The temperatures of 28 and 37 $^\circ\mathrm{C}$ correspond to the mosquito and host temperatures, respectively, whereas pH 7 and 5 correspond to the pH prior to cellular infection and to that of the virus within the late endosome. As observed in Figure 3b,c (see also Fig. S1, for all conditions), the characteristic angle of the intermediate state is, on average, approximately 55° and 52° in DENV-2 and DENV-3, respectively. Compared to the initial state, this corresponds to a reduction of approximately 22° and 25° for DENV-2 and DENV-3, respectively.
We then examined in detail the influence of temperature and pH on this angle. When raising the temperature from 28 to 37 $^\circ\mathrm{C}$ at pH 5, we observed a reduction in the angle of the intermediate state, whereas the opposite trend occurs at pH 7, in both serotypes (Fig. 3b,c). However, these changes are minimal and fall within the confidence interval. When raising the pH from 5 to 7, we observed a reduction in this angle at 28 $^\circ\mathrm{C}$, but the opposite trend was observed at 37 $^\circ\mathrm{C}$, in both serotypes. Here again, these differences are minimal and lie within the confidence interval. Taken together, these results indicate that, although pH and temperature slightly modulate the characteristic angle, the asymmetric state itself is observed against these variations.
This intermediate state, which we term the asymmetric state, resembles the compensatory embrace mechanism described in our previous work \cite{roponpalacios2026}. It is characterized by a reduction in the tilt angle relative to the initial state, which leads the pulled monomer to tilt toward the pulling axis (Z+ axis), as can be seen in Figures 3d,e. In addition, this state shows the exposure of the FL of the anchored monomer, while the contacts on the opposite side (cradle–FL) remain engaged until dissociation occurs (Fig. S2). Interestingly, this intermediate state exposes the immunodominant epitope of the FL \cite{A8} (Fig. 3d,e, see inset). This is particularly compelling, since stabilizing this intermediate state could facilitate recognition of the immunodominant FL epitope by a therapeutic agent, thereby blocking complete dissociation and the subsequent trimer formation.

\subsection*{Assessing the nature of native contacts in the DENV ecE and their dynamics during the dissociation} 

To understand the nature of the native contacts in DENV-2 and DENV-3, we analyzed the types of contact present at the interface of both dimers, comparing the three pH values (5, 6, and 7) for each serotype (Fig. 4a). We identified four types of interaction: van der Waals (vdW), which are the most abundant, followed by hydrogen bonds (hb) and, with comparable numbers to one another, salt bridges (sb) and $\pi$-cation interactions (pc). When comparing these interaction types from the initial structure at different pH values, we noted an increase of 1 hb and 5 vdW interactions at pH 5 in DENV-3, whereas no change was observed in DENV-2. Comparing the two serotypes, DENV-3 displays a greater number of hb and slightly more sb, whereas DENV-2 displays slightly more $\pi$-cation interactions and considerably more vdW.
The remaining analyses in this section were conducted under the most biologically relevant conditions (28/37 $^\circ\mathrm{C}$ and pH 5/7; for the other conditions, see Fig. S3–S7). We first analyzed the temporal dynamics of the hydrogen bonds (Fig. 4b), which shows the average number of these contacts over the replicas against time. When evaluating the effect of pH and temperature in DENV-2, we observed that the loss of hb is more abrupt at 37 $^\circ\mathrm{C}$ than at 28 $^\circ\mathrm{C}$; nevertheless, at 37 $^\circ\mathrm{C}$ a slight increase in the number of hb occurs at pH 5 during the first 2 ns (Fig. 4b). In DENV-3, by contrast, the loss of hb is rapid at both temperatures (Fig. 4b), showing a slight resistance around 4 ns at 28 $^\circ\mathrm{C}$, but a more linear decay at 37 $^\circ\mathrm{C}$. As for pH, there are no clear differences between the curves; only a subtle, more linear trend is observed at pH 7 than at pH 5, both at 37 $^\circ\mathrm{C}$.

 \begin{figure}[H]
        \includegraphics[width=1\textwidth]{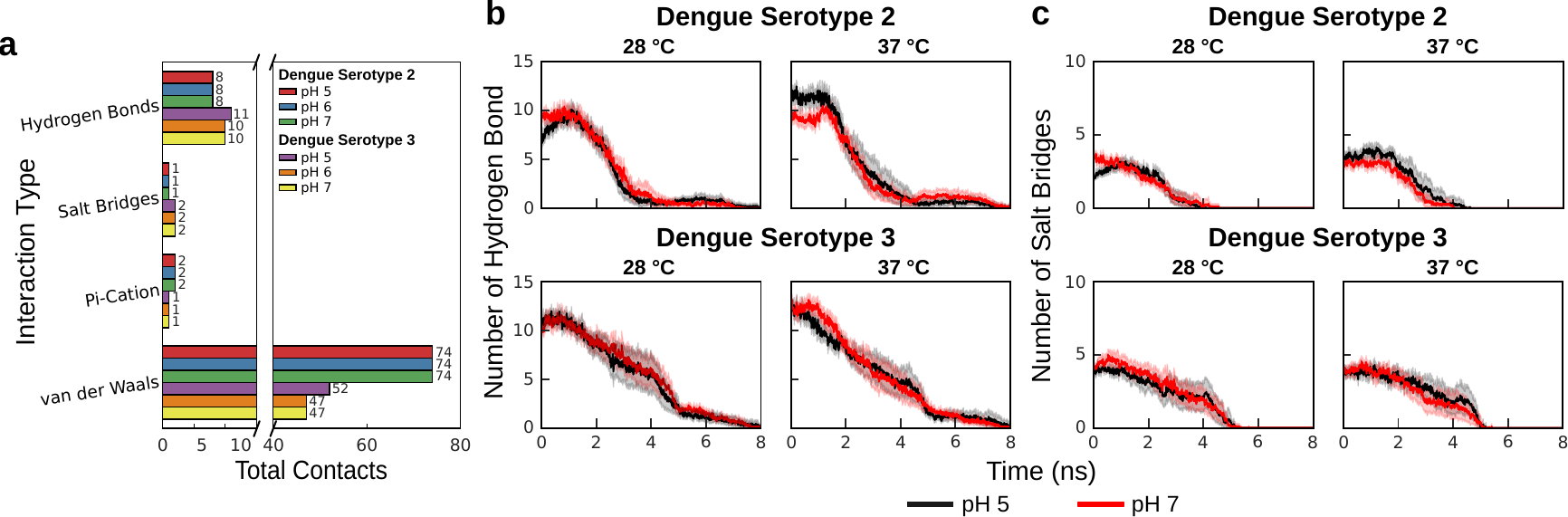}
    \centering
    \caption{\textbf{Noncovalent interactions at the DENV-2 and DENV-3 dimer interface.} (a) Intermonomeric contacts by interaction type under different pH conditions from the starting structure. (b) Intermonomeric hydrogen bonds and (c) salt bridges as a function of simulation time. Black and red curves correspond to pH 5 and pH 7, respectively. Lines indicate replica averages and shaded regions denote 95\% confidence intervals.
    \label{fig:Fig_4}}
\end{figure}

We next evaluated the salt bridges (Fig. 4c) and the effect of pH and temperature on them. In DENV-2, we observed differences upon varying the temperature: a subtle rapid decay at pH 5 and a minimal reduction at pH 7, both when going from 28 to 37 $^\circ\mathrm{C}$. As for pH, at 28 $^\circ\mathrm{C}$ the differences appear minimal, whereas at 37 $^\circ\mathrm{C}$ a slight increase in contacts is observed at pH 5 compared to pH 7. In DENV-3, the temperature increase slightly accentuated the linear decay at both pH values, whereas the pH values show minimal differences at both temperatures.
To gain further insight into the effect of pH and temperature on the interface interactions (Fig. 5), we analyzed the occupancy percentage of the residue pairs. For the hb in DENV-2, the comparison between temperatures (Fig. 5a, pairs in bold) revealed unique pairs, such as C105$_{\text{EDII}}$ – K310$_{\text{EDIII}}$ at 28 $^\circ\mathrm{C}$ and E314$_{\text{EDIII}}$ – G106$_{\text{EDII}}$ and E311$_{\text{EDIII}}$ - W101$_{\text{EDII}}$ at 37 $^\circ\mathrm{C}$. In DENV-3, this same comparison revealed Q269$_{\text{EDII}}$ – E247$_{\text{EDII}}$ at 37 $^\circ\mathrm{C}$. Regarding the effect of pH, we observed that, under acidic conditions, the occupancy percentage of several pairs (Fig. 5a,b, green) was reduced relative to pH 7.
The same analysis was performed for the salt bridges, in which no unique pairs were observed between temperatures for either serotype. When evaluating the effect of pH, we also noted pairs perturbed by acidification (Fig. 5c, green); overall, DENV-3 showed more perturbed pairs than DENV-2.
This perturbation, that is, the reduction in occupancy percentage at acidic pH, is particularly compelling, since reinforcing these pairs at neutral pH, or preventing them from being perturbed at acidic pH by means of some ligand, could hinder dimer dissociation in the endosome, thereby preventing fusion and the completion of infection.

 \begin{figure}[H]
        \includegraphics[width=1\textwidth]{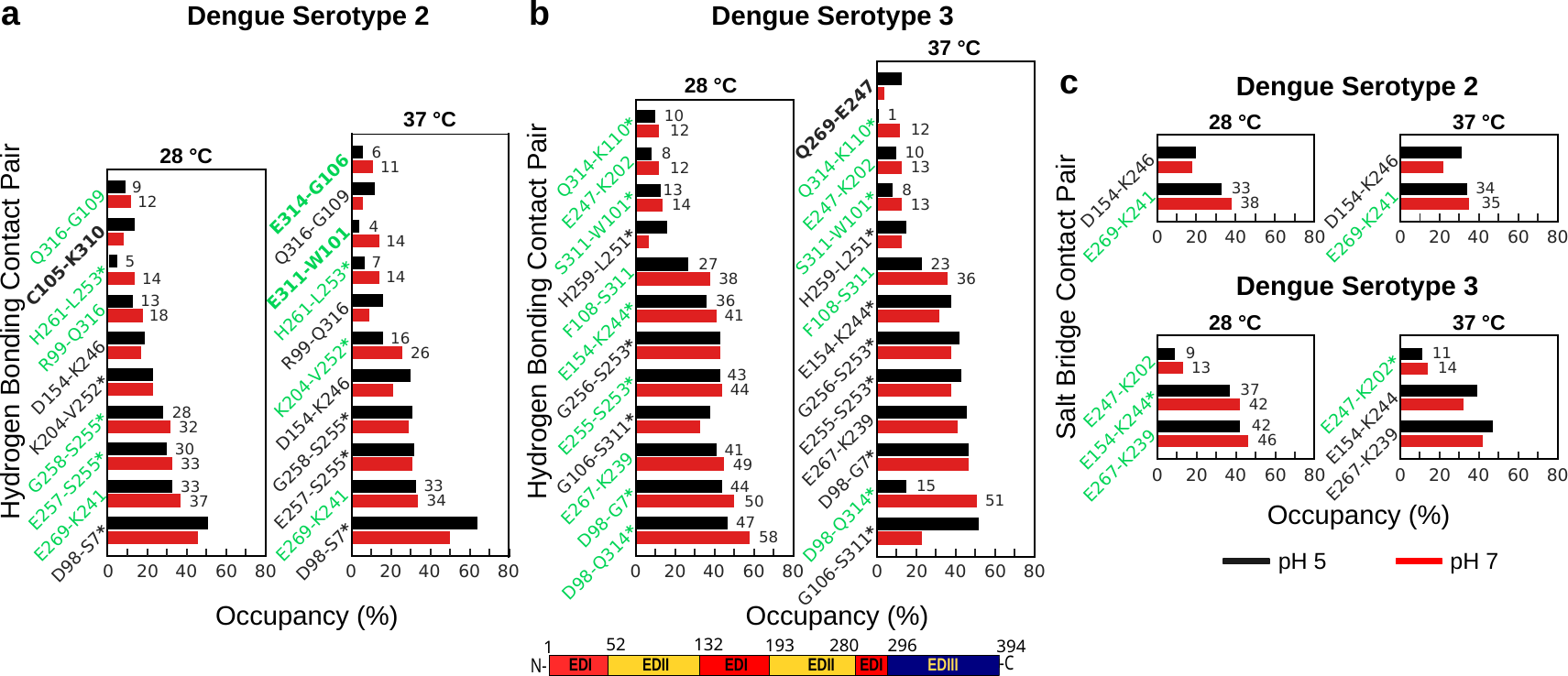}
    \centering
    \caption{\textbf{Intermonomeric contact occupancies during the dissociation of DENV ecE dimers. (a,b) Hydrogen bonds in DENV-2 and DENV-3, respectively. (c) Salt bridges in both serotypes.} Occupancies were calculated over 8 ns of simulation and combined across 20 independent replicas for pH 5 (black) and pH 7 (red) at 28 and 37 °C. Only contacts with a maximum occupancy $>$10\% in at least one condition are shown. Green labels indicate interactions with reduced occupancy at pH 5, and asterisks and bold text denote native contacts present in the initial crystal structures and pairs of contacts unique to the respective condition, respectively.
    \label{fig:Fig_5}}
\end{figure}

\section{DISCUSSION}
Dengue virus is a pathogen of major relevance to global public health \cite{A1, A11}. Its life cycle depends on the dissociation of the ecE and its subsequent interaction with the endosomal membrane \cite{A7}. This makes the protein an attractive therapeutic target \cite{A16}. In this work, we investigated the influence of pH and temperature on the mechanostability and dissociation process of the ecE dimer by means of SMD.
We first characterized the mechanostability of the interface from the force-extension curves, which proved comparable in both serotypes, with forces exceeding 1000 pN. These values surpass those observed for protein-receptor interfaces, such as SARS-CoV-2 spike:ACE2 (25-400 pN \cite{P8}) and the hyperstable Xmod-Doc:CohE complex (600-750 pN \cite{P11}), although they are lower than those of ultrastable interactions, such as Bbp:Fg$\alpha$ (3510 pN \cite{P7}) and SdrG:Fg$\beta$ complex, which exhibits force resilience comparable to that of covalent bonds \cite{P3}. We attribute this overestimation to the high pulling velocity (10 \text{\AA} ns$^{-1}$) employed in this work. Although this velocity increases the absolute forces observed, it was kept constant across all conditions, so that the relative comparisons remain valid.
In this comparison, we did not observe a clear effect of pH on mechanostability, since most differences remained within the confidence interval. This result is compatible with the literature, considering that pH acts primarily on the energetics of the conformational rearrangements \cite{A14, A7, A12, A15} rather than directly on the mechanical resistance of the interface. Indeed, analysis of the protonation states (Fig. 1) revealed that only a few interfacial histidines are protonated at acidic pH (H149 in both serotypes and, additionally, H259 in DENV-3 in an asymmetric manner), whereas at pH 6 and 7 no interfacial histidine is protonated. This reduced number of protonated histidines is consistent with the absence of a marked effect of pH on mechanical resistance. As for temperature, the effect was also subtle. Nevertheless, in DENV-3 we observed a reduction in the maximum force peak with increasing temperature, a behavior absent in DENV-2. This trend is consistent with the greater susceptibility of DENV-3 to temperature-induced monomerization reported by Kudlacek and collaborators \cite{A20}. Differences between serotypes were also evident in the dissociation profile, with DENV-2 showing a monotonic loss of contacts and DENV-3 exhibiting a second rupture point. Considering that the two differ by approximately 30–35\% in their sequence \cite{A8}, these observations suggest distinct mechanical properties of the interface.
Despite these serotype differences, the dissociation process shared a common feature across all analyzed conditions: it occurred asymmetrically. When quantified through the tilt angle and the cradle–FL distance, the two-dimensional distribution revealed the presence of an intermediate state resembling the compensatory embrace described previously \cite{roponpalacios2026}, characterized by a reduction in the tilt angle of approximately 22° in DENV-2 and 25° in DENV-3 relative to the initial state.
This intermediate possesses two properties that make it particularly attractive from a therapeutic standpoint. First, it represents a point along the dissociation pathway that, if stabilized or trapped, could prevent progression toward complete dimer separation and, consequently, toward the formation of the fusogenic trimer. In addition, in this same state the FL, an immunodominant epitope recognized by antibodies that, although poorly neutralizing, can become potently neutralizing following affinity maturation \cite{A8, A19}, is exposed, which could favor its recognition by antibodies or other molecules. Notably, this state proved robust to variations in pH and temperature, occurring even under neutral pH and physiological temperature conditions. This suggests that it could be populated and potentially captured even before viral entry into the cell or its arrival at the endosome. Nevertheless, this remains a hypothesis derived from the geometry of the intermediate observed in SMD, whose biological relevance requires experimental validation.
Finally, to understand which interactions sustain this interface, we characterized its composition. We observed that it is predominantly stabilized by vdW interactions, followed by hydrogen bonds, salt bridges, and pi-cation contacts. DENV-3 displayed a greater contribution from hydrogen bonds and salt bridges, whereas DENV-2 relies comparatively more on vdW interactions. These differences may be related to the distinct thermal sensitivity observed between serotypes, since the highly polar and charged DENV2 interface has reduced affinity as the temperature increases, in contrast to interactions driven by the hydrophobic effect, as discussed by Kudlacek and collaborators \cite{A20}.
Pairwise occupancy analysis identified contacts whose frequency decreases at acidic pH, with a greater number of perturbed pairs in DENV-3 than in DENV-2, in agreement with the role of acidification as a trigger for dissociation \cite{A14, A7}. \color{black}{These pH-sensitive pairs represent targets of interest for the development of molecules capable of reinforcing contacts that weaken in acidic environments, potentially increasing the energetic barrier and delaying the structural events required for viral fusion. It should be noted, however, that these observations derive from SMD simulations under static protonation, and their validation will require both experimental confirmation and simulations with dynamic protonation (cpHMD).}

\section{CONCLUSIONS}
In this work, we used SMD simulations to investigate the effects of pH and temperature on the mechanostability and dissociation of DENV-2 and DENV-3 ecE dimers. Both serotypes exhibited highly stable interfaces, with rupture forces above 1000 pN, although distinct dissociation profiles and thermal responses revealed serotype-specific mechanical behavior.
Dissociation consistently proceeded through an asymmetric pathway involving a metastable intermediate characterized by partial dimer opening and fusion loop exposure. Stabilization of this intermediate could, in principle, hinder complete dimer separation and subsequent formation of the fusogenic trimer, while its exposed regions may provide opportunities for molecular recognition.
Interfacial interactions were dominated by van der Waals contacts, and pairwise occupancy analysis identified pH-sensitive contacts, particularly in DENV-3, that may contribute to early dimer destabilization. These mechanistic and therapeutic implications should be further evaluated experimentally and through constant-pH molecular dynamics simulations.

\section*{ACKNOWLEDGMENTS}
This work was supported by Brazilian agencies: the National Council for Scientific and Technological Development (CNPq, grant \#409272/2021-3), 
the Coordination for the Improvement of Higher Education Personnel (CAPES,
\#88887.286296/2026-00), and the São Paulo Research Foundation (FAPESP, grants \#2022/00347-0, \#2025/02641-0 and \#2025/23025-6). Computational resources were provided by the National Laboratory for Scientific Computing (LNCC/MCTI, Brazil) through the SDumont supercomputer (Project antimicmd2, http://sdumont.lncc.br), and the EuroHPC Joint Undertaking for awarding access to the EuroHPC supercomputer LEONARDO, hosted by CINECA (Italy) through an EuroHPC Regular Access call. The authors acknowledge the use of large language models (LLMs) as AI-assisted tools during manuscript preparation for language editing and to support the debugging and documentation of analysis scripts. All AI-assisted outputs were reviewed by the authors, who take full responsibility for the final content of the manuscript.

\section*{AUTHOR CONTRIBUTIONS}
\textbf{G. R. P.}: conceived the study, developed the methodology and software, performed the simulations and formal analysis, curated the data, wrote the original draft, and prepared all figures.
\textbf{L.G.F.C}:  contributed to methodology and software development, performed simulations and formal analysis, curated data, wrote sections of the original draft, and prepared figures.
\textbf{J.R.}: contributed to the conceptualization and methodology, supervised the enhanced sampling calculations, and reviewed the manuscript.
\textbf{W.R.}: reviewed and edited the manuscript.
\textbf{A.S.A.}: conceived and supervised the study, administered the project, acquired funding, and reviewed the manuscript. All authors have read, reviewed, and approved the final version of the manuscript.

\section*{COMPETING INTERESTS}
The authors declare no conflict of interest.

\section*{DATA AND SOFTWARE AVAILABILITY}
In alignment with the FAIR (Findable, Accessible, Interoperable, and Reusable) principles for biomolecular simulations \cite{M18}, all essential simulation input files, output trajectories, and custom analysis scripts supporting this work will be openly accessible in a Zenodo repository upon publication.

\section*{SUPPORTING INFORMATION}

Supporting Information is available as supplementary material accompanying this preprint. Supplementary Figures S1–S7 characterize the conformational and interactional changes associated with the dissociation of DENV-2 and DENV-3 ecE dimers. Figure S1 presents normalized conformational landscapes across all simulated pH and temperature conditions. Figure S2 shows representative structural snapshots of intermediate and final expanded conformations, together with the cradle–FL separation distance. Figures S3–S7 describe the time evolution and occupancy profiles of intermonomeric hydrogen bonds and salt bridges, including native and transient contacts, across the 20 independent replicas.

\newpage

\bibliographystyle{achemso}
\renewcommand{\refname}{REFERENCES}
\bibliography{references}
\setcounter{section}{0}
\setcounter{equation}{0} \setcounter{figure}{0} \setcounter{table}{0}
\setcounter{page}{1} \makeatletter
\renewcommand{\thesection}{S\arabic{section}}
\renewcommand{\theequation}{S\arabic{equation}}
\renewcommand{\thefigure}{S\arabic{figure}}
\renewcommand{\bibnumfmt}[1]{[S#1]}
\renewcommand{\citenumfont}[1]{S#1}

\clearpage
\nolinenumbers
\normalsize
\section*{SUPPORTING INFORMATION}

The Supporting Information is available free of charge at the ACS Publications website.

\begin{figure}[H]
    \includegraphics[width=1\textwidth]{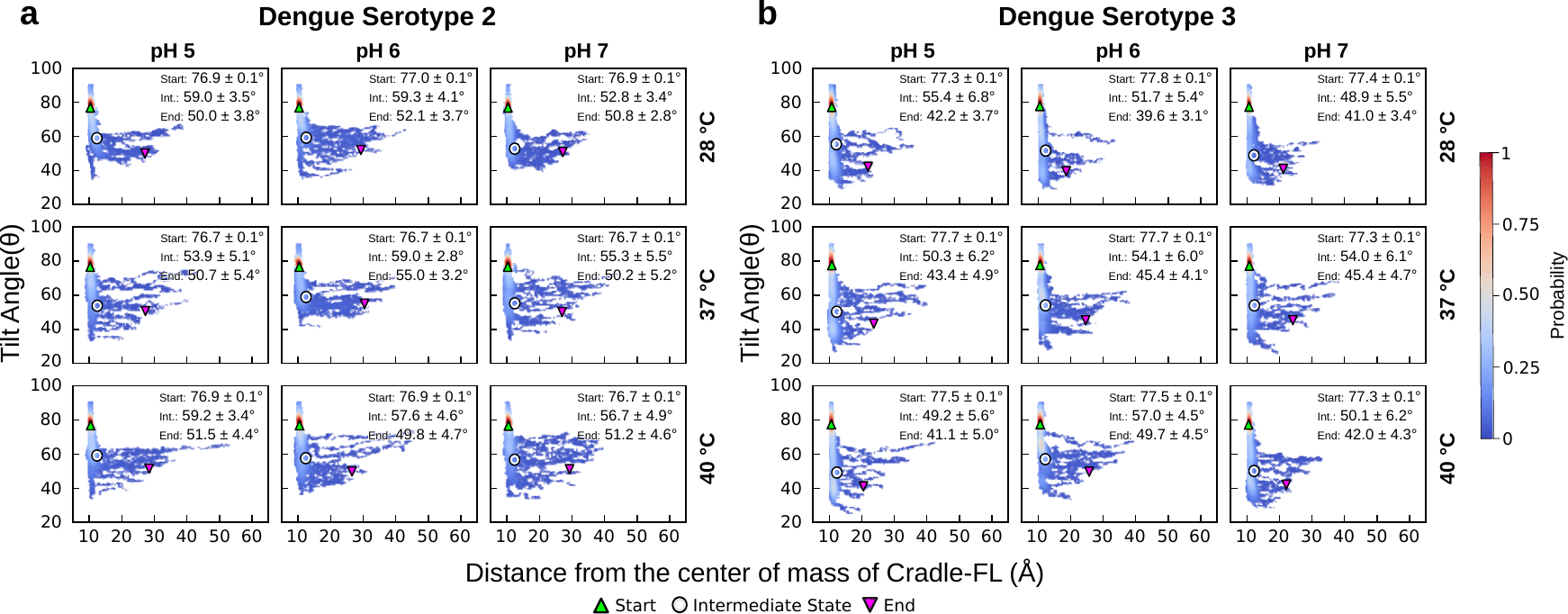}
    \centering
    \caption*{\textbf{Figure S1.} Conformational landscapes of DENV-2 and DENV-3 ecE dimers during dissociation. Normalized probability density maps for (a) DENV-2 and (b) DENV-3 across all simulated conditions of pH (5, 6, and 7) and temperature (28, 37, and 40 $^\circ\mathrm{C}$). The maps show the structural transition from the Start to the Intermediate (Int.) and End states. Symbols indicate the averages from 20 independent replicas.
    \label{fig:Figure_S1}}
\end{figure}
\vspace{1cm}
\begin{figure}[H]
    \includegraphics[width=1\textwidth]{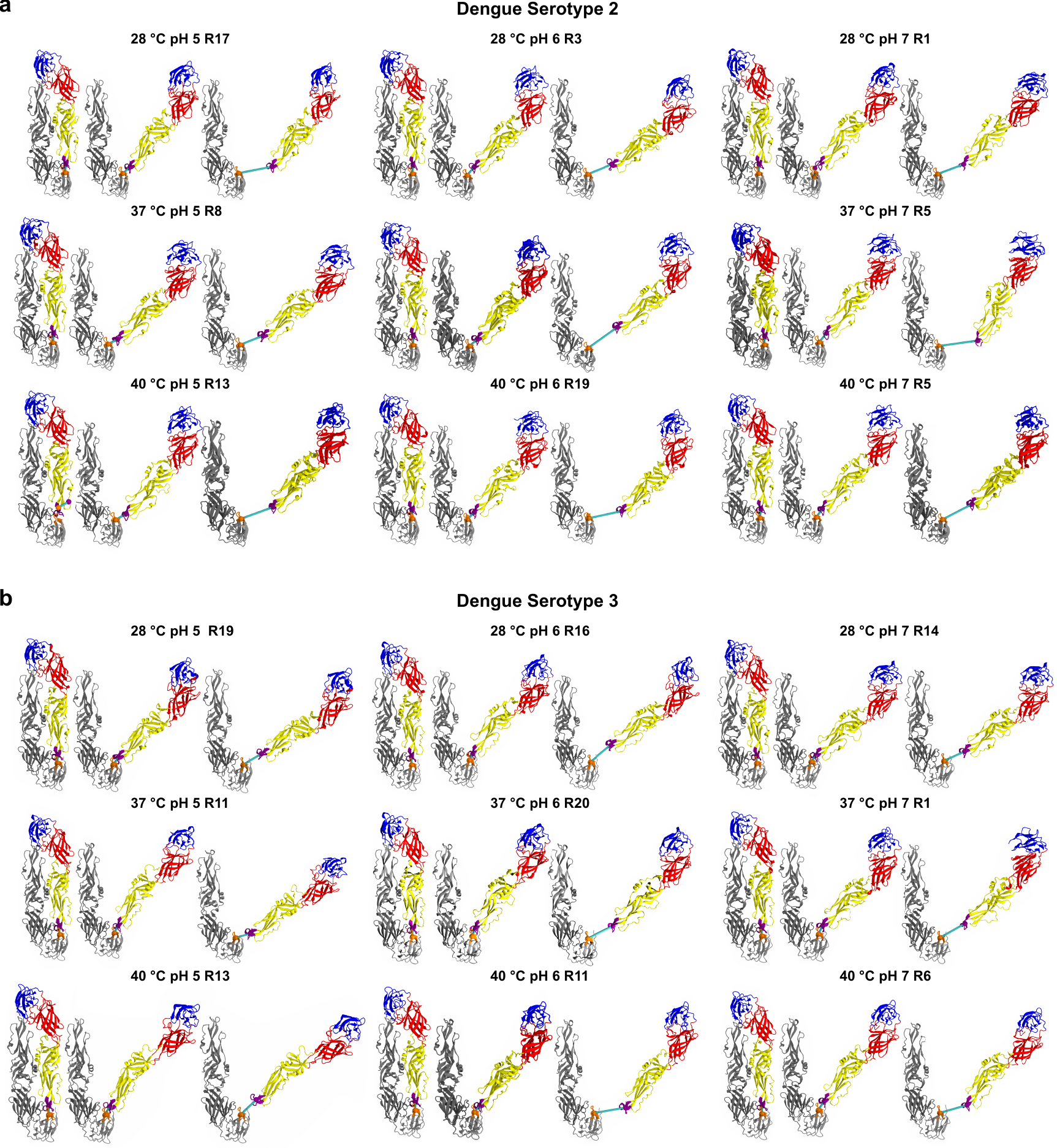}
    \centering
    \caption*{\textbf{Figure S2.} Comprehensive structural snapshots of ecE dimer dissociation for (a) DENV-2 and (b) DENV-3 across all simulated temperatures and pH conditions. For each tested condition, a representative replica (e.g., R17) is shown, depicting the dimer at the intermediate state (left) and the final expanded conformation (right). The cyan line illustrates the separation distance between the cradle–FL.
    \label{fig:Figure_S2}}
\end{figure}
\vspace{1cm}
\begin{figure}[H]
    \includegraphics[width=1\textwidth]{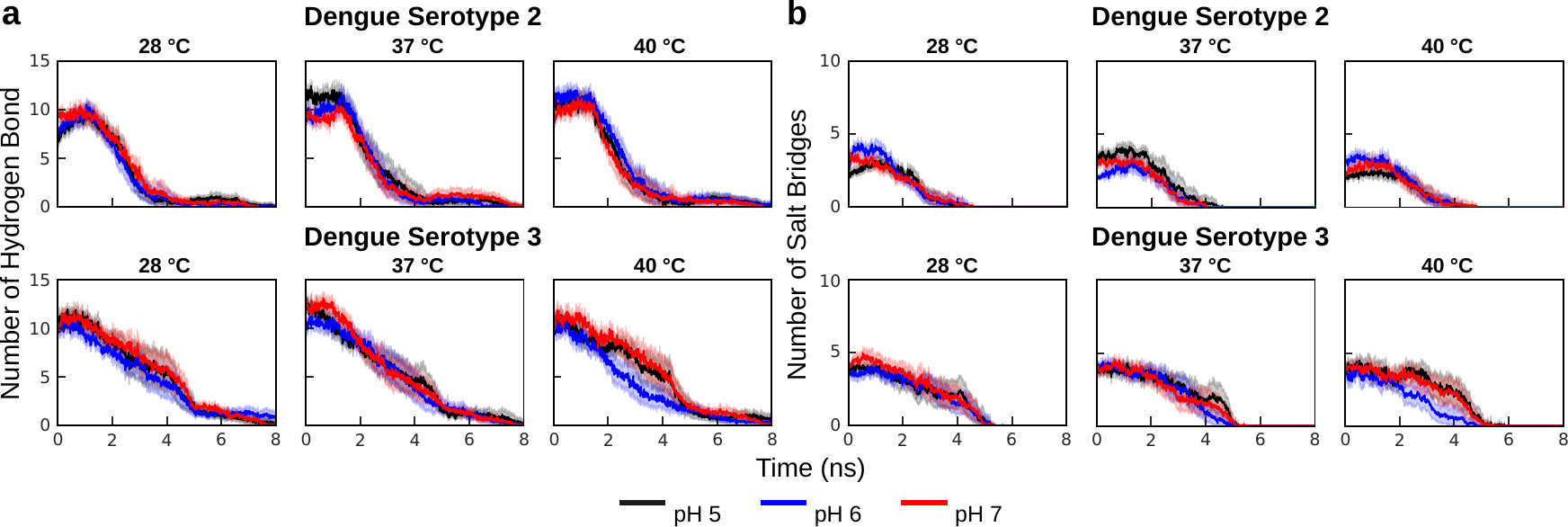}
    \centering
    \caption*{\textbf{Figure S3.} Time evolution of noncovalent interactions at the DENV-2 and DENV-3 dimer interface. (a) Intermonomeric hydrogen bonds and (b) salt bridges as a function of simulation time across all evaluated temperatures (28, 37, and 40 $^\circ\mathrm{C}$). Black, blue, and red curves correspond to pH 5, pH 6, and pH 7, respectively. Lines indicate replica averages and shaded regions denote 95\% confidence intervals.}
    \label{fig:Figure_S3}
\end{figure}
\vspace{1cm}
\begin{figure}[H]
    \includegraphics[width=0.70\textwidth]{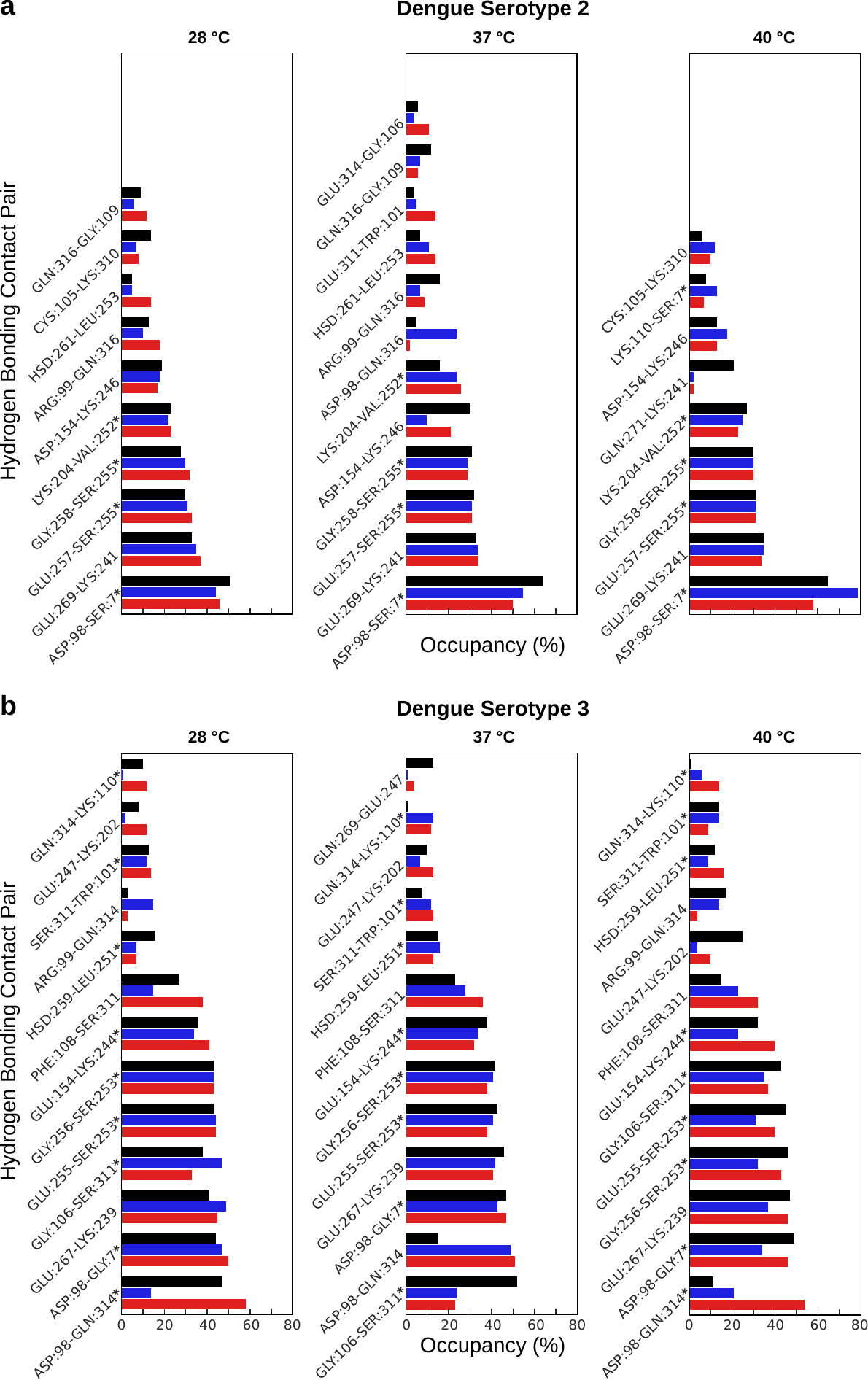}
    \centering
    \caption*{{\textbf{Figure S4.} Intermonomeric hydrogen bond occupancies during the dissociation of DENV ecE dimers. Profiles for (a) DENV-2 and (b) DENV-3 are shown. Occupancies were calculated over 8 ns of simulation and combined across 20 independent replicas for pH 5 (black), pH 6 (blue), and pH 7 (red) at 28, 37, and 40 $^\circ\mathrm{C}$. Only contacts with a maximum occupancy $>$10\% in at least one condition are shown. Asterisks denote native contacts present in the initial crystal structures.}
    \label{fig:Figure_S4}}
\end{figure}
\vspace{1cm}
\begin{figure}[H]
    \includegraphics[width=0.75\textwidth]{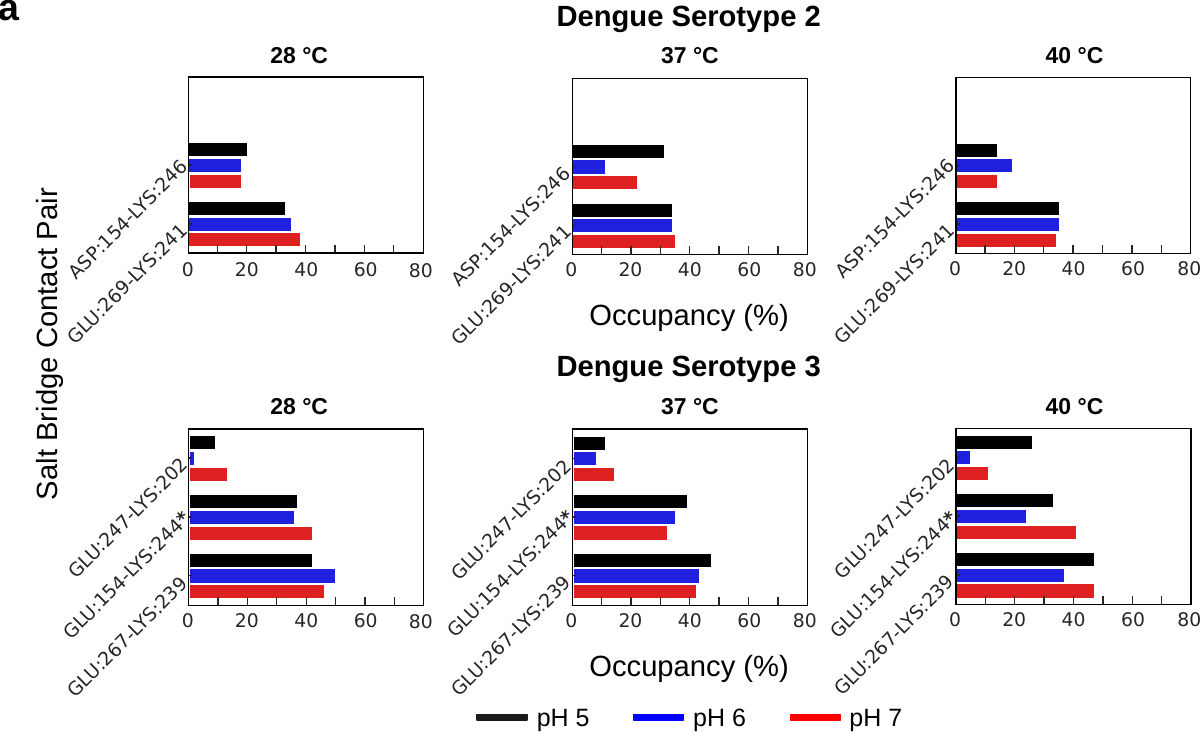}
    \centering
    \caption*{{\textbf{Figure S5.} Intermonomeric salt bridge occupancies during the dissociation of DENV ecE dimers. Occupancies were calculated over 8 ns of simulation and combined across 20 independent replicas for pH 5 (black), pH 6 (blue), and pH 7 (red) at 28, 37, and 40 $^\circ\mathrm{C}$. Only contacts with a maximum occupancy $>$10\% in at least one condition are shown. Asterisks denote native contacts present in the initial crystal structures.}
    \label{fig:Figure_S5}}
\end{figure}
\vspace{1cm}
\begin{figure}[H]
    \includegraphics[width=0.8\textwidth]{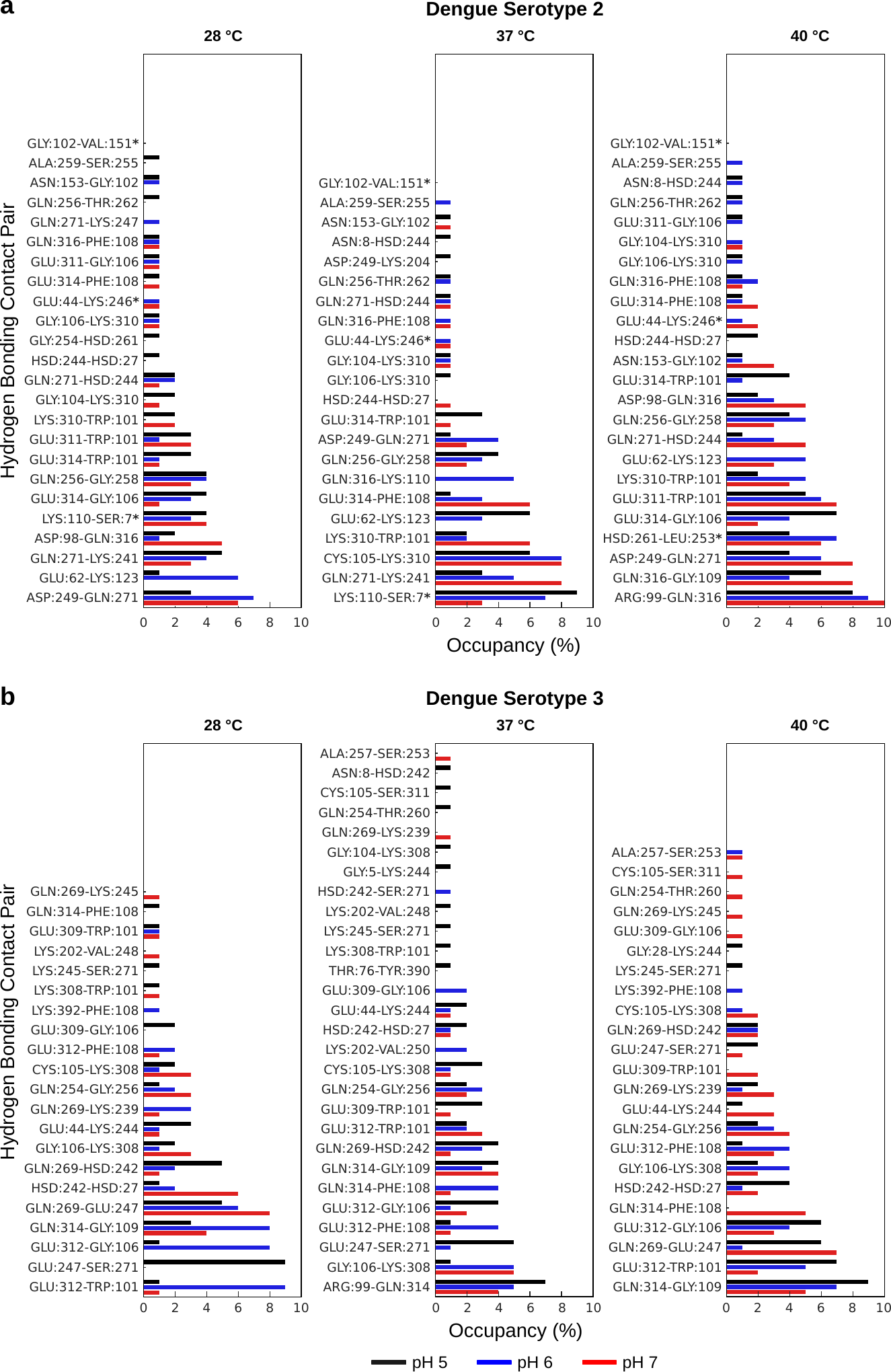}
    \centering
    \caption*{{\textbf{Figure S6.} Transient intermonomeric hydrogen bond occupancies during the dissociation of DENV ecE dimers. Profiles for (a) DENV-2 and (b) DENV-3 are shown. Occupancies were calculated over 8 ns of simulation and combined across 20 independent replicas for pH 5 (black), pH 6 (blue), and pH 7 (red) at 28, 37, and 40 $^\circ\mathrm{C}$. Only transient contacts with a maximum occupancy $<$10\% across all evaluated conditions are shown. Asterisks denote native contacts present in the initial crystal structures.}
    \label{fig:Figure_S6}}
\end{figure}
\vspace{1cm}
\begin{figure}[H]
    \includegraphics[width=0.75\textwidth]{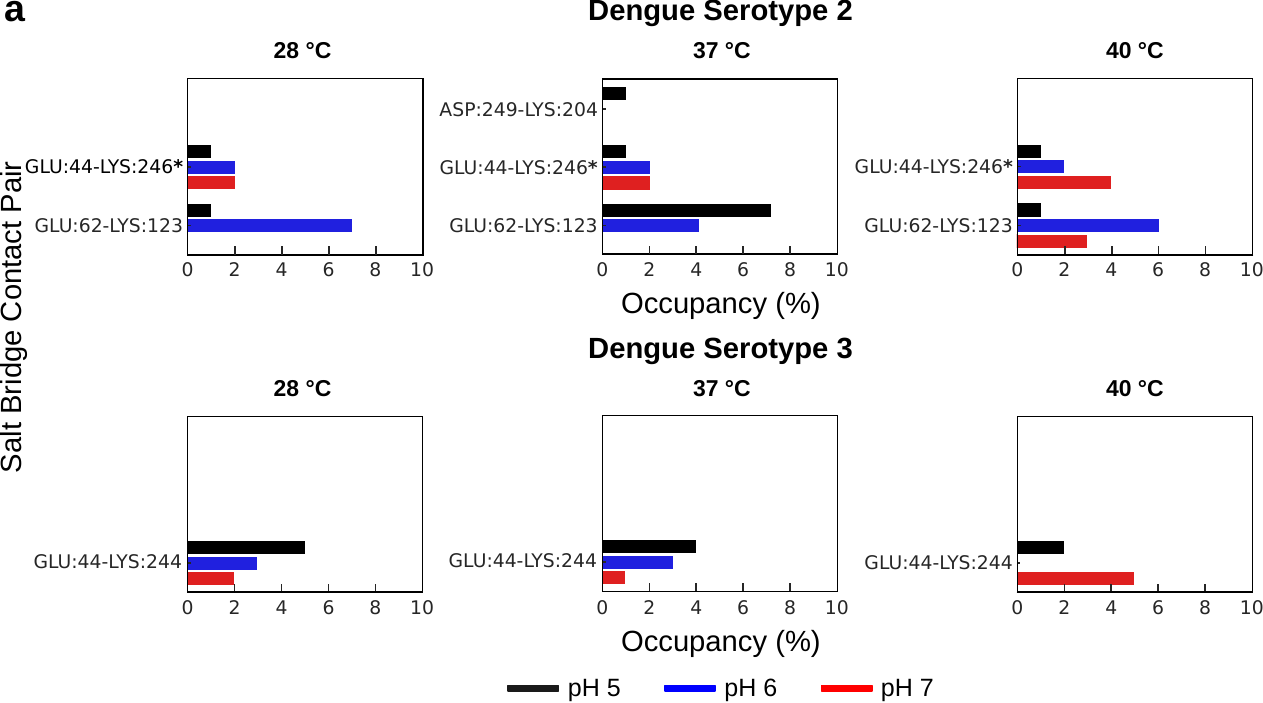}
    \centering
    \caption*{{\textbf{Figure S7.} Transient intermonomeric salt bridge occupancies during the dissociation of DENV ecE dimers. Profiles for (a) DENV-2 and DENV-3 are shown. Occupancies were calculated over 8 ns of simulation and combined across 20 independent replicas for pH 5 (black), pH 6 (blue), and pH 7 (red) at 28, 37, and 40 $^\circ\mathrm{C}$. Only transient contacts with a maximum occupancy $<$10\% across all evaluated conditions are shown. Asterisks denote native contacts present in the initial crystal structures.}
    \label{fig:Figure_S7}}
\end{figure}

\end{document}